\documentclass[12pt]{SelfArx} % Document font size and equations flushed left
\usepackage[english]{babel} % Specify a different language here - english by default
\usepackage{amsfonts,amssymb,amsbsy,amsmath,amsthm}
\usepackage{listings}
\usepackage{lscape}
\usepackage{multirow}
\usepackage{threeparttable}
\usepackage{here}
\usepackage{url}
\usepackage[numbers]{natbib}

\theoremstyle{definition}

\theoremstyle{plain}

\definecolor{color1}{RGB}{0,0,90} % Color of the article title and sections
\definecolor{color2}{RGB}{0,20,20} % Color of the boxes behind the abstract and headings

\usepackage{hyperref} % Required for hyperlinks
\hypersetup{hidelinks,colorlinks,breaklinks=true,urlcolor=red,citecolor=color1,linkcolor=color1,bookmarksopen=false,
	pdftitle={pimeta: an R package of prediction intervals for random-effects meta-analysis},
	pdfauthor={Kengo Nagashima, Hisashi Noma, Toshi A. Furukawa}}

\JournalInfo{Preprint (14-Aug-2019)} % Journal information
\Archive{ } % Additional notes (e.g. copyright, DOI, review/research article)

\PaperTitle{pimeta: an R package of prediction intervals for random-effects meta-analysis}
\ShortTitle{R package of prediction intervals for random-effects meta-analysis}

\Authors{Kengo Nagashima\textsuperscript{1}*, Hisashi Noma\textsuperscript{2}, and Toshi A. Furukawa\textsuperscript{3}} % Authors
\affiliation{\footnotesize{\textsuperscript{1}\textit{Research Center for Medical and Health Data Science, The Institute of Statistical Mathematics, Japan.}}}
\affiliation{\footnotesize{\textsuperscript{2}\textit{Department of Data Science, The Institute of Statistical Mathematics, Japan.}}}
\affiliation{\footnotesize{\textsuperscript{3}\textit{Departments of Health Promotion and Human Behavior, Kyoto University Graduate School of Medicine/School of Public Health, Kyoto, Japan.}}}
\affiliation{\footnotesize{*\textbf{Corresponding author}: nshi1201@gmail.com}}

 % Keywords - if you don't want any simply remove all the text between the curly brackets
\newcommand{\keywordname}{Keywords} % Defines the keywords heading name

\Abstract{
\footnotesize{
The prediction interval is gaining prominence in meta-analysis as it enables the assessment of uncertainties in treatment effects and heterogeneity between studies.
However, coverage probabilities of the current standard method for constructing prediction intervals cannot retain their nominal levels in general, particularly when the number of synthesized studies is moderate or small, because their validities depend on large sample approximations.
Recently, several methods have developed been to address this issue.
This paper briefly summarizes the recent developments in methods of prediction intervals and provides readable examples using R for multiple types of data with simple code.
The pimeta package is an R package that provides these improved methods to calculate accurate prediction intervals and graphical tools to illustrate these results.
The pimeta package is listed in ``CRAN Task View: Meta-Analysis.''
The analysis is easily performed in R using a series of R packages.
}
}

\begin{document}

\flushbottom % Makes all text pages the same height

\maketitle % Print the title and abstract box

%\tableofcontents % Print the contents section

\thispagestyle{empty} % Removes page numbering from the first page

%----------------------------------------------------------------------------------------
%	ARTICLE CONTENTS
%----------------------------------------------------------------------------------------

\section{Introduction}\label{sec:intro}
Random-effects meta-analyses combine the treatment effect estimates across studies, accounting for heterogeneity.
One of the objectives of a random-effects meta-analysis is to summarize studies and to make statistical inferences on the average treatment effect.
Quantification and evaluation of the magnitude of heterogeneity are also very important, because true treatment effects can differ for each study due to differences in patient characteristics, follow-up times, drug regimens, and other factors, in many cases.
However, researchers often interpret results from random-effects models in the same manner as those from fixed-effect models that assume the true effect does not differ from study to study \cite{RileyGates2011,RileyHiggins2011}.
It means that they tend to pay little attention to the treatment effects in each study that can be different from the average treatment effect.

Skipka \cite{Skipka2006} first proposed a prediction interval in random-effects meta-analysis, although, in which the term heterogeneity interval was used.
Aterwards, it was recommended that a prediction interval \cite{Higgins2009} should be reported alongside a confidence interval and heterogeneity measures \cite{RileyHiggins2011,Veroniki2019}.
It can be interpreted as the range of the predicted true treatment effect in a new study, given information of the combined studies.
Prediction intervals provide useful additional information to confidence intervals, because a prediction interval is a measure of treatment effects that accounts for heterogeneity.
The importance of prediction intervals has come to be recognized.
Several methods to construct prediction intervals for random-effects meta-analysis have been developed.

Higgins et al. \cite{Higgins2009} proposed a method to calculate the prediction interval that can provide valuable information from a very simple formula.
This prediction interval is useful because it can also be calculated from summarized results (i.e., the standard error of the average effect and estimate of heterogeneity parameter) of a random-effects meta-analysis.
However, Partlett \& Riley \cite{Partlett2017} confirmed that the Higgins et al.'s prediction interval could have poor coverage when the heterogeneity or the number of studies is small, because this prediction interval is based on large-sample approximations.

Partlett \& Riley \cite{Partlett2017} proposed modified methods of the Higgins et al.'s prediction interval using restricted maximum likelihood (REML) estimation with four variance estimators for the overall mean effect as follows:
(1) the approximate variance estimator;
(2) the Hartung--Knapp variance estimator \cite{Hartung1999,Hartung2001};
(3) the Sidik--Jonkman bias-corrected variance estimator \cite{Sidik2006};
and (4) the Kenward--Roger's approach \cite{Kenward1997}.
These prediction intervals exhibit better coverage performances than Higgins et al.'s prediction interval, but are still insufficient when the heterogeneity or number of studies is small \cite{Partlett2017}, because these methods are also based on similar approximations of the Higgins et al.'s prediction interval.

To address these issues, Nagashima et al. \cite{Nagashima2018} proposed an alternative prediction interval based on a parametric bootstrap method using a confidence distribution \cite{Schweder2002,Xie2013}.
In their simulation studies, the Nagashima et al.'s prediction interval has been found to perform very well even the heterogeneity and number of studies are small; this prediction interval was accurate when $I^2$-statistics were in the range of $7\%$--$99\%$ and no. of studies $\geq 3$.

Nagashima et al. \cite{pimeta2018} have developed the R \cite{R2019} package pimeta to facilitate implementation of these improved methods in meta-analyses.
The pimeta package is available from the Comprehensive R Archive Network (CRAN) at \url{https://CRAN.R-project.org/package=pimeta} or GitHub repository \url{https://github.com/nshi-stat/pimeta}.
Moreover, the pimeta package is listed in ``CRAN Task View: Meta-Analysis'' \cite{CRANTaskView2019}.

The remainder of this paper is organized as follows.
Section \ref{sec:prediction} briefly reviews the random-effects model for meta-analysis and prediction intervals.
Section \ref{sec:pimeta} introduces the analysis of prediction intervals using R with the pimeta package and provides examples based on meta-analysis datasets included in the package.
Section \ref{sec:conclusion} concludes the paper.

%% -- Manuscript -----------------------------------------------------------
\section{Prediction intervals for random-effects meta-analysis}\label{sec:prediction}

\subsection{Random-effects model}\label{sec:random}
Let us consider the case of combining information from a series of $K$ comparative studies by the random-effects model \cite{Cochran1937,DerSimonian1986}.
The random-effects model can be defined as follows:
\begin{equation} \label{eq:randomeffects}
	\begin{array}{lll}
		Y_k &=& \theta_k+\epsilon_k,\\
		\theta_k &=& \mu+u_k,
	\end{array}
\end{equation}
where $Y_k$ is a random variable of an effect size estimate from the $k$th study, $\theta_k$ is the true treatment effect in the $k$th study, $\epsilon_k$ is a random error within a study, $\mu$ is an unknown parameter of the overall mean effect, $u_k$ is a random variable reflecting study-specific deviation from the overall mean effect, and $\epsilon_k$ and $u_k$ are assumed to be mutually independent and normally distributed, which means that $\epsilon_k \sim N(0, \sigma_k^2)$, $u_k \sim N(0, \tau^2)$, $\sigma_k^2>0$, $\tau^2>0$.
The parameters of the within-study variances, $\sigma_k^2$, are known and replaced by their efficient estimates \cite{Biggerstaff2008}.
The parameter of the heterogeneity variance is an unknown parameter that should be estimated.

The overall mean effect, $\mu$, can be estimated as $\hat{\mu}=(\sum_{k=1}^K \hat{w}_k Y_k)/(\sum_{k=1}^K \hat{w}_k)$, where $\hat{w}_k=(\sigma_k + \hat{\tau})^{-1}$, and $\hat{\tau}$ is an estimator of the heterogeneity variance that has been proposed by a number of researchers \cite{Sidik2007,Veroniki2016,Petropoulou2017,Langan2018} (see also Section \ref{sec:citau}).
Similarly, many methods to estimate a confidence interval of the overall mean effect, $\mu$, have also been proposed \cite{Brockwell2001,Meca2008,Veroniki2019} (see also Section \ref{sec:citau}).

\subsection{Higgins--Thompson--Spiegelhalter prediction interval}\label{sec:hts}
Higgins et al. \cite{Higgins2009} proposed a simple plug-in type prediction interval that can be expressed as
\begin{equation} \label{eq:hts}
	\left[
	\hat{\mu}-t_{K-2}^{\alpha}\sqrt{\hat{\tau}_{DL}^2+\widehat{\mathrm{Var}}[\hat{\mu}]},~
	\hat{\mu}+t_{K-2}^{\alpha}\sqrt{\hat{\tau}_{DL}^2+\widehat{\mathrm{Var}}[\hat{\mu}]}
	\right],
\end{equation}
where $\widehat{\mathrm{Var}}[\hat{\mu}]=1/(\sum_{k=1}^K \hat{w}_k)$ is an approximate variance estimator of $\hat{\mu}$, $t_{K-2}^{\alpha}$ is the $100(1-\alpha/2)$ percentile of the $t$ distribution with $K-2$ degrees of freedom, $\hat{\tau}_{DL}^2$ is the DerSimonian--Laird estimator of the heterogeneity variance \cite{DerSimonian1986} that is defined as $\hat{\tau}_{DL}^2=\max[0, \{Q-(K-1)\}/\{S_1+S_2/S_1\}]$, $Q=\sum_{k=1}^K v_k(Y_k-\bar{Y})^2$, $S_r=\sum_{k=1}^K v_k^r$, $v_k=(\sigma_k^{2})^{-1}$, and $\bar{Y}=(\sum_{k=1}^K v_k Y_k) / (\sum_{k=1}^K v_k)$.
This prediction interval is essentially based on the following two approximations: $(\hat{\mu}-\mu)/\sqrt{\widehat{\mathrm{Var}}[\hat{\mu}]}$ is approximately distributed as $N(0, 1)$, and $(K-2)(\hat{\tau}_{DL}^2+\widehat{\mathrm{Var}}[\hat{\mu}])/(\tau^2+\mathrm{Var}[\hat{\mu}])$ is approximately distributed as $\chi^2(K-2)$.
These approximations generally do not hold for small $K$.
Therefore, this prediction interval perform well when the heterogeneity and number of studies are moderate \cite{Partlett2017}.
This prediction interval was accurate when $I^2$ were in the range of $42\%$--$99\%$ and $K > 25$ in simulations \cite{Nagashima2018}.
However, in practice, most meta-analyses include less than 20 studies \cite{Kontopantelis2013}.

\subsection{Partlett--Riley prediction intervals}\label{sec:pr}
Partlett \& Riley \cite{Partlett2017} proposed prediction intervals following restricted maximum likelihood (REML) estimation of the heterogeneity variance \cite{Sidik2007,Harville1977} with four variance estimators for the overall mean effect, $\mu$.
This prediction interval can be expressed as
\[
\left[
\hat{\mu}_R-t_{K-2}^{\alpha}\sqrt{\hat{\tau}_{R}^2+\widehat{\mathrm{Var}}[\hat{\mu}_R]},~
\hat{\mu}_R+t_{K-2}^{\alpha}\sqrt{\hat{\tau}_{R}^2+\widehat{\mathrm{Var}}[\hat{\mu}_R]}
\right].
\]

\begin{enumerate}
	\item The approximate variance estimator.\\
	This prediction interval can be expressed as
	\[
	\left[
	\hat{\mu}_R-t_{K-2}^{\alpha}\sqrt{\hat{\tau}_{R}^2+\widehat{\mathrm{Var}}[\hat{\mu}_R]},~
	\hat{\mu}_R+t_{K-2}^{\alpha}\sqrt{\hat{\tau}_{R}^2+\widehat{\mathrm{Var}}[\hat{\mu}_R]}
	\right].
	\]
	This method replaces $\hat{\mu}$, $\hat{\tau}_{DL}^2$, and $\widehat{\mathrm{Var}}[\hat{\mu}]$ in the Higgins et al.'s prediction interval (\ref{eq:hts}) with $\hat{\mu}_R$, $\hat{\tau}_{R}^2$, and $\widehat{\mathrm{Var}}[\hat{\mu}_R]$, where $\hat{\tau}_{R}^2$ is an iterative solution of the equation
	\[
	\hat{\tau}_{R}^2=
	\frac{\sum_{k=1}^K \hat{w}_{R,k}^2\{(Y_k - \hat{\mu}_{R})^2 + 1/\sum_{l=1}^K \hat{w}_{R,k} - \sigma_k^2\}}{\sum_{k=1}^K \hat{w}_{R,k}^2}
	,
	\]
	$\hat{w}_{R,k}=(\sigma_k^2+\hat{\tau}_R^2)^{-1}$, and $\hat{\mu}_{R}=\sum_{k=1}^K \hat{w}_{R,k} Y_k / \sum_{k=1}^K \hat{w}_{R,k}$.
	\item The Hartung--Knapp variance estimator \cite{Hartung1999,Hartung2001}.\\
	As the above one, this method replaces $\hat{\mu}$, $\hat{\tau}_{DL}^2$, and $\widehat{\mathrm{Var}}[\hat{\mu}]^2$ in (\ref{eq:hts}) with $\hat{\mu}_R$, $\hat{\tau}_{R}^2$, and $\widehat{\mathrm{Var}}_{HK}[\hat{\mu}_R]$, where the Hartung--Knapp estimator is defined as
	\[
	\widehat{\mathrm{Var}}_{HK}[\hat{\mu}_{R}]=
	\frac{1}{K-1}\sum_{k=1}^K \frac{\hat{w}_{R,k}(Y_k-\hat{\mu}_{R})^2}{\sum_{l=1}^K \hat{w}_{R,l}}.
	\]
	\item The Sidik--Jonkman bias-corrected variance estimator \cite{Sidik2006}.\\
	Similarly, this method replaces $\hat{\mu}$, $\hat{\tau}_{DL}^2$, and $\widehat{\mathrm{Var}}[\hat{\mu}]$ in (\ref{eq:hts}) with $\hat{\mu}_R$, $\hat{\tau}_{R}^2$, and $\widehat{\mathrm{Var}}_{SJ}[\hat{\mu}_{R}]$, where the Sidik--Jonkman bias-corrected estimator
	\[
	\widehat{\mathrm{Var}}_{SJ}[\hat{\mu}_{R}]=
	\frac{\sum_{k=1}^K \hat{w}_{R,k}^2 (1-\hat{h}_k)^{-1} (Y_k-\hat{\mu}_{R})^2}{(\sum_{k=1}^K \hat{w}_{R,k})^2}
	,
	\]
	and
	\[
	\hat{h}_k=
	\frac{2\hat{w}_{R,k}}{\sum_{l=1}^K\hat{w}_{R,l}}-
	\frac{\sum_{l=1}^K \hat{w}_{R,l}^2(\sigma_l^2+\hat{\tau}_{R}^2)}{(\sigma_k^2+\hat{\tau}_{R}^2) \sum_{l=1}^K \hat{w}_{R,l}^2}.
	\]
	\item The Kenward--Roger's approach \cite{Kenward1997,Morris2018}.\\
	This prediction interval can be expressed as
	\[
	\left[
	\hat{\mu}_R-t_{\nu-1}^{\alpha}\sqrt{\hat{\tau}_{R}^2+\widehat{\mathrm{Var}}_{KR}[\hat{\mu}_R]},~
	\hat{\mu}_R+t_{\nu-1}^{\alpha}\sqrt{\hat{\tau}_{R}^2+\widehat{\mathrm{Var}}_{KR}[\hat{\mu}_R]}
	\right].
	\]
	This method applies the approximate degrees of freedom, $\nu-1$, and a bias-adjusted variance estimator, $\widehat{\mathrm{Var}}_{KR}[\hat{\mu}_R]$.
	For the random-effects model in (\ref{eq:randomeffects}), the expected information for $\tau^2$ is given by
	\[
	\mathcal{I}(\hat{\tau}_R^2)=
	\frac{1}{2}\sum_{k=1}^K \hat{w}_{R,k}^2-
	\frac{\sum_{k=1}^K \hat{w}_{R,k}^3}{\sum_{k=1}^K \hat{w}_{R,k}^2}+
	\frac{1}{2}\left(\frac{\sum_{k=1}^K \hat{w}_{R,k}^2}{\sum_{k=1}^K \hat{w}_{R,k}}\right).
	\]
	Thus, the bias-adjusted variance estimator is defined as
	\[
	\widehat{\mathrm{Var}}_{KR}[\hat{\mu}_{R}]=
	\frac{1}{\sum_{k=1}^K \hat{w}_{R,k}}+\frac{2\left\{ \frac{\sum_{k=1}^K \hat{w}_{R,k}^3}{\sum_{k=1}^K \hat{w}_{R,k}}- \left(\frac{\sum_{k=1}^K \hat{w}_{R,k}^2}{\sum_{k=1}^K \hat{w}_{R,k}}\right)^2 \right\}}{\mathcal{I}(\hat{\tau}_R^2) \sum_{k=1}^K \hat{w}_{R,k}},
	\]
	and the approximate degrees of freedom is $\nu=2\mathcal{I}(\hat{\tau}_R^2)/(\widehat{\mathrm{Var}}_{KR}[\hat{\mu}_R] \sum_{k=1}^K \hat{w}_{R,k}^2)^2$.
\end{enumerate}
These prediction intervals are refinements of the Higgins et al.'s prediction interval to provide more appropriate intervals.
However, these prediction intervals are based on similar approximations of the Higgins et al.'s prediction interval.
They were accurate when $I^2$ were in the range of $40\%$--$99\%$ and $K > 25$ in simulations \cite{Nagashima2018}.

\subsection{Nagashima--Noma--Furukawa prediction interval}\label{sec:nnf}
Nagashima et al. \cite{Nagashima2018} proposed a prediction interval that is valid under the heterogeneity variance with a small number of studies.
To avoid using the approximations of the Higgins et al.'s prediction interval, this prediction interval is based on a parametric bootstrap approach using a confidence distribution \cite{Schweder2002,Xie2013} to account for the uncertainty of $\hat{\tau}_{DL}^2$ with an exact distribution estimator of $\tau^2$ and the Hartung--Knapp variance estimator \cite{Hartung1999,Hartung2001}.

This prediction interval can be constructed using the algorithm described below:
\begin{enumerate}
	\item Generate $B$ bootstrap samples $\tilde{\tau}_b^2$ ($b=1, \ldots, B$) that are drawn from the exact distribution of $\hat{\tau}_{UDL}^2=\{Q-(K-1)\}/\{S_1+S_2/S_1\}$, $z_b$ that are drawn from $N(0, 1)$, and $t_b$ that are drawn from $t(K-1)$.
	\item Calculate $\tilde{\mu}_b=\sum_{k=1}^K \tilde{w}_{bk}Y_k/\sum_{k=1}^K \tilde{w}_{bk}$, and $\tilde{\theta}_{new,b}=\tilde{\mu}_b+z_b \tilde{\tau}_b-t_b\sqrt{\widetilde{\mathrm{Var}}_{HK}[\tilde{\mu}_b]}$, where $\tilde{w}_{bk}=(\sigma_k^2+\tilde{\tau}_b^2)^{-1}$, and  $\widetilde{\mathrm{Var}}_{HK}[\tilde{\mu}_b]=\frac{1}{K-1} \sum_{k=1}^K \frac{\tilde{w}_{bk}(Y_k-\tilde{\mu}_b)^2}{\sum_{l=1}^K \tilde{w}_{bl}}$.
	\item Calculate the prediction limits $c_l$ and $c_u$ that are $100\times \alpha/2$ and $100\times(1-\alpha/2)$ percentage points of $\tilde{\theta}_{new,b}$, respectively.
\end{enumerate}
The exact distribution function of the random variable $\hat{\tau}_{UDL}^2$ is given by
\[
H(\tilde{\tau}^2)=\Pr(\hat{\tau}_{UDL}^2<\tilde{\tau}^2)=1-F_Q(q_{obs}; \tilde{\tau}^2),
\]
where $F_Q$ is the distribution function of $Q$, and $q_{obs}$ is an observed value of $Q$.
By applying the inverse transformation method, a random sample $\tilde{\tau}_b^2=H^{-1}(u_b)$ can be computed by numerical inversion of $H(\tilde{\tau}_b^2)=u_b$, where $u_b$ is a random number with  a uniform distribution, $U(0, 1)$.
If $H(0) > u$, then the sample is truncated to zero ($\tilde{\tau}_b^2=0$).
A quadratic form, $Q$, has the same distribution as the random variable $\sum_{k=1}^K \lambda_k \chi^2_k(1)$ that has only one parameter $\tau^2$, where $\lambda_k \geq 0$ are the eigenvalues of matrix $\mathbf{\mathrm{S}}$, $\chi_1^2(1), \chi_2^2(1), \ldots, \chi_K^2(1)$ are $K$ independent central chi-square random variables, each with one degree of freedom, $\mathbf{\mathrm{S}}=\mathbf{\mathrm{\Sigma}}^{1/2}\mathbf{\mathrm{A}}\mathbf{\mathrm{\Sigma}}^{1/2}$, $\mathbf{\mathrm{\Sigma}}=\mathrm{diag}(\sigma_1^2+\tau^2, \sigma_2^2+\tau^2, \ldots, \sigma_K^2+\tau^2)$, $\mathbf{\mathrm{A}}=\mathbf{\mathrm{V}} - \mathbf{\mathrm{v}}\mathbf{\mathrm{v}}^{\mathrm{T}}/v_+$, $\mathbf{\mathrm{V}}=\mathrm{diag}(v_1, v_2, \ldots, v_K)$, $\mathbf{\mathrm{v}}=(v_1, v_2, \ldots, v_K)^{\mathrm{T}}$, and $v_+ =\sum_{k=1}^K v_k$ \cite{Biggerstaff2008}.

This prediction interval has been found to perform very well even the heterogeneity is small and number of studies is small; this prediction interval was accurate when $I^2$ were in the range of $7\%$--$99\%$ and $K \geq 3$ in simulation studies \cite{Nagashima2018}.

\section{Calculating prediction intervals using R} \label{sec:pimeta}

\subsection{The pimeta package} \label{sec:pima}

The prediction intervals introduced in Section \ref{sec:prediction} are implemented in the pimeta package.
One of the main functions is `pima()', and its simple usage is:
\begin{verbatim}
	library("pimeta")
	sbp <- data.frame(
	y = c(0.00, 0.10, -0.40, -0.80, -0.63, 0.22, -0.34, -0.51, 0.03, -0.81),
	sigmak = c(0.42347717, 0.21939179, 0.02551067, 0.19898325, 0.30102594,
	0.30102594, 0.07142988, 0.10204269, 0.12245123, 0.30102594),
	label = c("Almond (2001)", "Cashew (2003)", "Pecan (2004)",
	"Macadamia (2007)", "Pistachio (2008)", "Hazelnut (2011)",
	"Coconut (2012)", "Walnut (2014)", "Chestnut (2015)",
	"Peanut (2017)")
	)
	piboot <- pima(sbp$y, sbp$sigmak, seed = 3141592)
	print(piboot)
\end{verbatim}

The output is:
\begin{verbatim}
	Prediction & Confidence Intervals for Random-Effects Meta-Analysis
	
	A parametric bootstrap prediction and confidence intervals
	Heterogeneity variance: DerSimonian-Laird
	Variance for average treatment effect: Hartung (Hartung-Knapp)
	
	No. of studies: 10
	
	Average treatment effect [95% prediction interval]:
	-0.3341 [-0.8789, 0.2165]
	d.f.: 9
	
	Average treatment effect [95% confidence interval]:
	-0.3341 [-0.5673, -0.0985]
	d.f.: 9
	
	Heterogeneity measure
	tau-squared: 0.0282
	I-squared:  70.5%
\end{verbatim}

Here, `y' is the vector of effect size estimates $(Y_1, Y_2, \ldots, Y_K)$, `se' is the vector of parameters of the within-study standard errors $(\sigma_1, \sigma_2, \ldots, \sigma_K)$, and `method' is the name of the calculation method for a prediction interval (the default is ``boot'' that corresponds to the Nagashima et al.'s prediction interval [see Section \ref{sec:nnf}]).
Because the Nagashima et al.'s prediction interval uses bootstrapping, the function `pima()' provides the `parallel' option to reduce the computational time required for bootstrapping.
Other arguments for the `pima' function are shown in Table \ref{tab:pima}.

%% Table 1
\begin{table}[p]
	\centering
	\caption{\label{tab:pima} Arguments for the `pima()' function.}
	\begin{tabular}{llp{9.5cm}} \hline
		Argument & Type & Description \\ \hline
		`y' & numeric vector & The vector of effect size estimates. \\
		`se' & numeric vector & The vector of within-study standard errors; `se' should be $>0$; either `se' or `v' must be specified. \\
		`v' & numeric vector & The vector of within-study variances; `v' should be $>0$; either `se' or `v' must be specified. \\
		`method' & character & The calculation method; `HTS` indicates Higgins et al., ``APX'', ``HK'', ``SJ'', or ``KR'' indicates Partlett--Riley, ``boot'' indicates Nagashima et al.; the default is ``boot''. \\
		`alpha' & numeric & The alpha level; the default is 0.05. \\
		`B' & integer & The number of bootstrap samples; `B' should be $>0$; the default is 25000; ignored if `method' is not ``boot''. \\
		`parallel' & integer/logical & The number of threads used in parallel computing or `FALSE'; the default is `FALSE', which means single threading; ignored if `method' is not ``boot''. \\
		`seed' & integer & Set the value of random seed; ignored if `method' is not ``boot''. \\
		`maxit1' & integer & The maximum number of iterations for the exact distribution function of $\hat{\tau}_{UDL}^2$; `maxit1' should be $>1$; the default is 10000; ignored if `method' is not ``boot''. \\
		`eps' & numeric & The desired level of accuracy for the exact distribution function of $\hat{\tau}_{UDL}^2$; `eps' should be $> 0$; the default is $10^{-10}$; ignored if `method' is not ``boot''. \\
		`lower' & numeric & The lower limit of random numbers of $\hat{\tau}_{UDL}^2$; `lower' should be $\geq 0$; the default is 0; ignored if `method' is not ``boot''. \\
		`upper' & numeric & The upper limit of random numbers of $\hat{\tau}_{UDL}^2$; `upper' should be $> 0$; the default is 1000; ignored if `method' is not ``boot''. \\
		`maxit2' & integer & The maximum number of iterations for numerical inversions; `maxit2' should be $> 0$ the default is 1000; ignored if `method' is not ``boot''. \\
		`tol' & numeric & The desired accuracy for numerical inversions; `tol' should be $> 0$; the default is `.Machine\$double.eps\^{0.25}; ignored if `method' is not ``boot''. \\
		`rnd' & numeric vector & A vector of random numbers from the exact distribution of $\hat{\tau}_{UDL}^2$; if `rnd' is specified, then the step of random numbers generations for $\hat{\tau}_{UDL}^2$ will be skipped; ignored if `method' is not ``boot''. \\
		`maxiter' & integer & The maximum number of iterations for REML estimation; `maxiter' should be $> 0$; the default is 100; ignored if `method' is not ``APX'', ``HK'', ``SJ'', or ``KR''. \\ \hline
	\end{tabular}
\end{table}

The `pima()' function can also estimate the Higgins et al.'s (see Section \ref{sec:hts}) and Partlett \& Riley's (see Section \ref{sec:pr}) prediction intervals by setting the method option.

\begin{verbatim}
	pima(sbp$y, sbp$sigmak, method = "HTS")
	pima(sbp$y, sbp$sigmak, method = "HK")
\end{verbatim}

\begin{figure}[ht]
\centering
\includegraphics{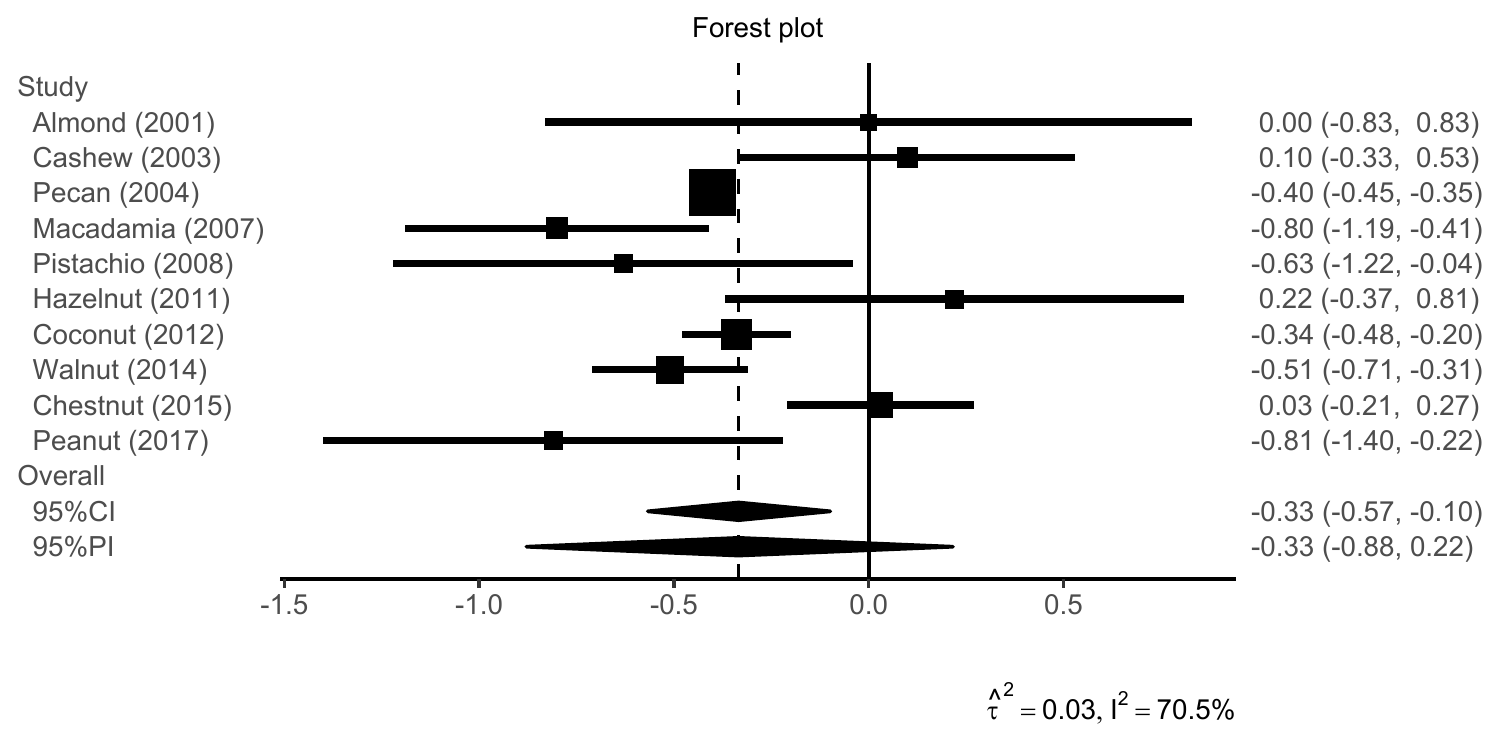}
\caption{\label{fig:forest} A forest plot for a `pima' object.}
\end{figure}

An estimated result can be summarized by the `print()' method for a `pima' object returned by the `pima()' function resulting in a list that is shown in Table \ref{tab:pimaout}.
Both prediction and confidence intervals, and summary statistics that are usually reported in a random-effects meta-analysis, are provided.
The method used in the confidence interval corresponds to that in the prediction interval.
An estimated result can also be summarized by a forest plot by the `plot()' method, as shown in Figure \ref{fig:forest}.
The forest plot function has an argument `studylabel' to set the labels for each study.

%% Table 2
\begin{table}[h]
	\centering
	\caption{\label{tab:pimaout} Output of the `pima()' function.}
	\begin{tabular}{ll} \hline
		Output & Description \\ \hline
		`K' & the number of studies, $K$ \\
		`muhat' & the average treatment effect estimate, $\hat{\mu}$ \\
		`lpi', `upi' & the lower and upper prediction limits \\
		`lci', `uci' & the lower and upper confidence limits \\
		`nup' & the degrees of freedom for the prediction interval \\
		`nuc' & the degrees of freedom for the confidence interval \\
		`tau2h' & the estimate for $\tau^2$ \\
		`i2h' & the estimate for $I^2$ \\ \hline
	\end{tabular}
\end{table}

%% Figure 1

\begin{verbatim}
	cairo_pdf("forestplot.pdf", width = 6, height = 3, family = "Arial")
	plot(piboot, digits = 2, base_size = 10)
	dev.off()
\end{verbatim}

\subsection{Random-effects meta-analysis with binary outcomes} \label{sec:illustrations_binom}

The pimeta package can handle random-effects meta-analysis of controlled clinical trials with binary outcomes.
The `convert\_bin()' function in the pimeta package is an implementation of methods described in Hartung \& Knapp \cite{Hartung2001b}.
Here, $m_{kj}$ is the number of successes, $n_{kj}$ is the number of patients, and $p_{kj}$ is the true binomial probability of success in treatment group $j$ (for $j=1,2$) in the $k$th study.
Logarithmic odds ratios are given by
\[
\log \widehat{\mathrm{OR}}_{k}=\log \left\{
\frac{m_{k1}+0.5}{(n_{k1}-m_{k1})+0.5}
\frac{(n_{k2}-m_{k2})+0.5}{m_{k2}+0.5}
\right\},
\]
and its variances are given by
\[
\widehat{\mathrm{Var}}[\widehat{\mathrm{OR}}_{k}]=\frac{1}{m_{k1}+0.5}+\frac{1}{(n_{k1}-m_{k1})+0.5}+\frac{1}{m_{k2}+0.5}+\frac{1}{(n_{k2}-m_{k2})+0.5}.
\]
Logarithmic relative risks are given by
\[
\log \widehat{\mathrm{RR}}_{k}=\log \frac{(m_{k1}+0.5)(n_{k2}+0.5)}{(n_{k1}+0.5)(m_{k2}+0.5)},
\]
and its variances are given by
\[
\widehat{\mathrm{Var}}[\widehat{\mathrm{RR}}_{k}]=\frac{1}{m_{k1}+0.5}-\frac{1}{n_{k1}+0.5}+\frac{1}{m_{k2}+0.5}-\frac{1}{n_{k2}+0.5}.
\]
Risk differences are given by
\[
\widehat{\mathrm{RD}}_{k}=\frac{m_{k1}}{n_{k1}}-\frac{m_{k2}}{n_{k2}},
\]
and its variances are given by
\[
\widehat{\mathrm{Var}}[\widehat{\mathrm{RD}}_{k}]=\widehat{\mathrm{Var}}[\hat{p}_{k1}]+\widehat{\mathrm{Var}}[\hat{p}_{k2}],~
\widehat{\mathrm{Var}}[\hat{p}_{kj}]=\frac{1}{n_{kj}}\left(\frac{m_{kj}+1/16}{n_{kj}+1/8}\right)\left\{\frac{(n_{kj}-m_{kj})+1/16}{n_{kj}+1/8}\right\}.
\]
These estimates are used as effect size estimates and within-study variances (e.g., $Y_k=\log \widehat{\mathrm{OR}}_{k}$ and $\sigma_k^2=\widehat{\mathrm{Var}}[\log \widehat{\mathrm{OR}}_{k}]$.)

The `convert\_bin()' function converts binary outcome data to effect size estimates and within-study standard error vectors.
This function has an argument `type' that is a character indicating an outcome measure (i.e., ``logOR'', ``logRR'', or ``RD'' can be specified).
Input data `m1', `n1', `m2', and `n2', which are integer vectors, are converted to effect size estimates $Y_k$ and within-study standard errors $\sigma_k$.
A dataset of 13 placebo-controlled trials with cisapride that was reported by Hartung \& Knapp \cite{Hartung2001b} was analyzed.
This dataset is reproduced in Table \ref{tab:cisapride}.

%% Table 3
\begin{table}[h]
	\centering
	\caption{\label{tab:cisapride} The cisapride data \citep{Hartung2001b}.}
	\begin{tabular}{ccccc} \hline
		Study & Treatment (cisapride) group & Placebo group \\
		& [m1/n1] & [m2/n2] \\ \hline
		Creytens & 15/16 & 9/16 \\
		Milo & 12/16 & 1/16 \\
		Francois and De Nutte & 29/34 & 18/34 \\
		Deruyttere et al. & 42/56 & 31/56 \\
		Hannon & 14/22 & 6/22 \\
		Roesch & 44/54 & 17/55 \\
		De Nutte et al. & 14/17 & 7/15 \\
		Hausken and Bestad & 29/58 & 23/58 \\
		Chung & 10/14 & 3/15 \\
		Van Outryve et al. & 17/26 & 6/27 \\
		Al-Quorain et al. & 38/44 & 12/45 \\
		Kellow et al. & 19/29 & 22/30 \\
		Yeoh et al. & 21/38 & 19/38 \\ \hline
	\end{tabular}
\end{table}

\begin{figure}[h]
\centering
\includegraphics{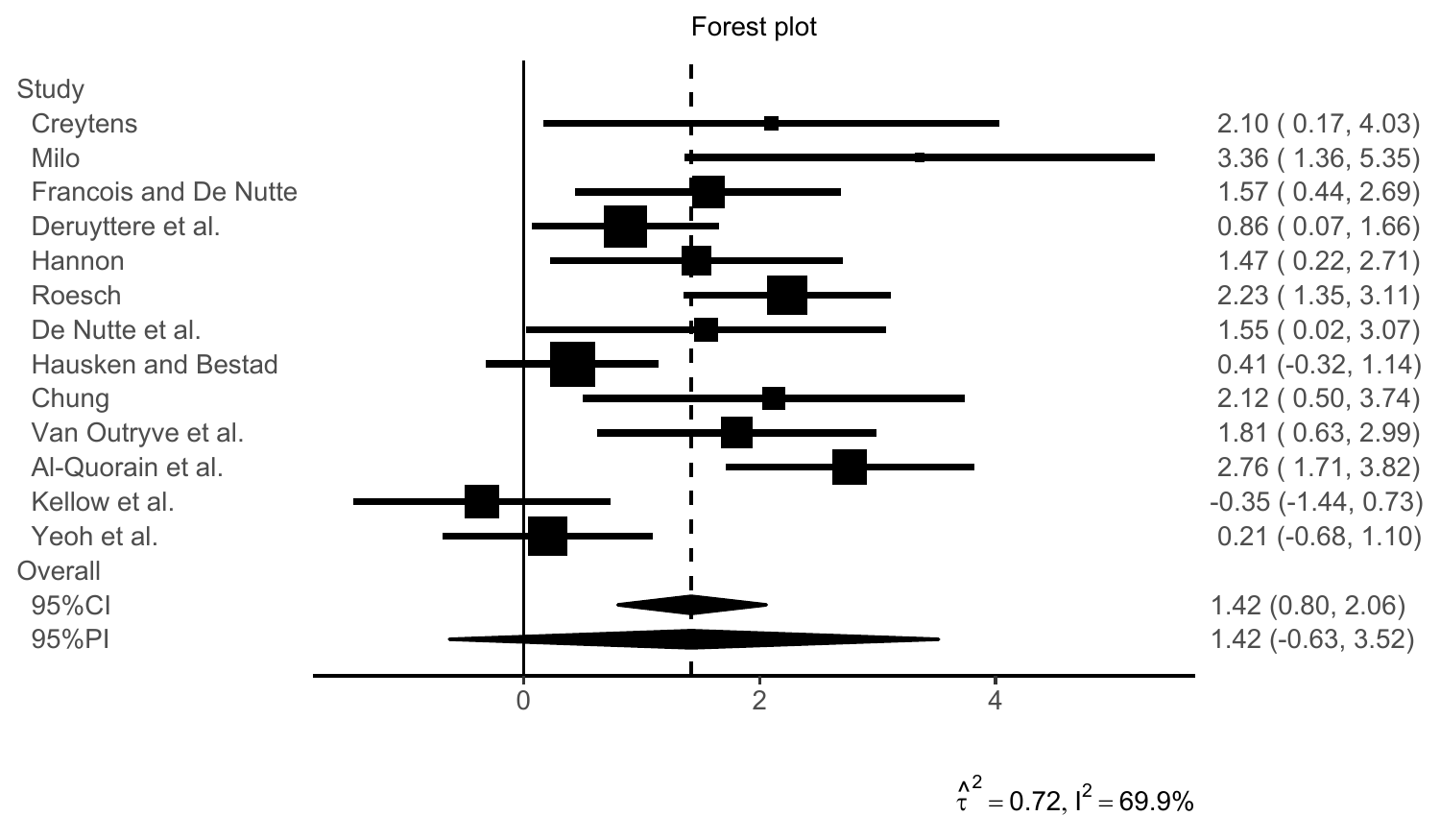}
\caption{\label{fig:forest_bin} A forest plot of the cisapride data.}
\end{figure}

The converted data can also be analyzed by the `pima()' function, and the results can be summarized as a forest plot (see Figure \ref{fig:forest_bin}).

%% Figure 2

\begin{verbatim}
	m1 <- c(15,12,29,42,14,44,14,29,10,17,38,19,21)
	n1 <- c(16,16,34,56,22,54,17,58,14,26,44,29,38)
	m2 <- c( 9, 1,18,31, 6,17, 7,23, 3, 6,12,22,19)
	n2 <- c(16,16,34,56,22,55,15,58,15,27,45,30,38)
	dat <- convert_bin(m1, n1, m2, n2, type = "logOR")
	pibin <- pima(dat$y, dat$se, seed = 2236067)
	print(pibin)
	binlabel <- c(
	"Creytens", "Milo", "Francois and De Nutte", "Deruyttere et al.",
	"Hannon", "Roesch", "De Nutte et al.", "Hausken and Bestad",
	"Chung", "Van Outryve et al.", "Al-Quorain et al.", "Kellow et al.",
	"Yeoh et al.")
	cairo_pdf("forestplot_bin.pdf", width = 6, height = 3.5, family = "Arial")
	plot(pibin, digits = 2, base_size = 10, studylabel = binlabel)
	dev.off()
\end{verbatim}

The coverage performances of prediction intervals have been discussed by Nagashima et al. \cite{Nagashima2018}.
The Nagashima et al.'s prediction interval achieved the nominal level of coverage under realistic meta-analysis settings.

\subsection{Confidence intervals and heterogeneity variance estimators} \label{sec:citau}

Because Riley \& Higgins \cite{RileyHiggins2011} recommended that a prediction interval should be reported alongside a confidence interval and heterogeneity measure, the pimeta package also provides functions to calculate confidence intervals for the overall mean effect and heterogeneity variance estimators.

The `cima()' function calculates a confidence interval.
The arguments of the `cima()' and `pima()' functions are identical except for the `method' as shown in Table \ref{tab:cima}.

%% Table 4
\begin{table}[h]
	\centering
	\caption{\label{tab:cima} The `method' option for the `cima()' function.}
	\begin{tabular}{lp{12cm}} \hline
		`method' & Description \\ \hline
		``boot'' & A parametric bootstrap confidence interval \citep{Nagashima2018}; the default is ``boot''. \\ 
		``DL'' & A Wald-type $t$-distribution confidence interval. The DerSimonian--Laird estimator for $\tau^2$ and an approximate variance estimator for the overall mean effect are used. \\
		``APX'' & A Wald-type $t$-distribution confidence interval. The REML estimator for $\tau^2$ and an approximate variance estimator are used. \\
		``HK'' & A Wald-type $t$-distribution confidence interval. The REML estimator for $\tau^2$ and the Hartung--Knapp variance estimator are used \citep{Hartung1999,Hartung2001}. \\
		``SJ'' & A Wald-type $t$-distribution confidence interval. The REML estimator for $\tau^2$ and the Sidik--Jonkman variance estimator are used \citep{Sidik2006}. \\
		``KR'' & A Wald-type $t$-distribution confidence interval. The REML estimator for $\tau^2$ and the Kenward--Roger approach are used \citep{Partlett2017}. \\ 
		``PL'' & Profile likelihood confidence interval \citep{Hardy1996}. \\ 
		``BC'' & Profile likelihood confidence interval with a Bartlett-type correction \citep{Noma2011}. \\ \hline
	\end{tabular}
\end{table}

For example, the following script calculates a profile likelihood confidence interval with a Bartlett-type correction.
\begin{verbatim}
	cima(sbp$y, sbp$sigmak, method = "BC")
\end{verbatim}

The `tau2h()' function calculates a heterogeneity variance estimate.
This function has arguments to specify methods for the heterogeneity variance estimator and its confidence interval; `method' and `methodci' as shown in Table \ref{tab:tau2h}.

%% Table 5
\begin{table}[h]
	\centering
	\caption{\label{tab:tau2h} The `method' and `methodci' options for the `tau2h()' function.}
	\begin{tabular}{lp{12cm}} \hline
		`method' & Description \\ \hline
		``DL'' & DerSimonian--Laird estimator \citep{DerSimonian1986}. \\ 
		``VC'' & Variance component type estimator \citep{Hedges1983}. \\
		``PM'' & Paule--Mandel estimator \citep{Paule1982}. \\
		``HM'' & Hartung--Makambi estimator \citep{Hartung2003}. \\
		``HS'' & Hunter--Schmidt estimator \citep{Hunter2004}.
		This estimator has a negative bias \citep{Viechtbauer2005}. \\
		``ML'' & Maximum likelihood (ML) estimator \citep{DerSimonian1986}. \\
		``REML'' & Restricted maximum likelihood (REML) estimator \citep{DerSimonian1986}. \\
		``AREML'' & Approximate restricted maximum likelihood estimator \citep{Thompson1999}. \\
		``SJ'' & Sidik--Jonkman estimator \citep{Sidik2005}. \\
		``SJ2'' & Sidik--Jonkman improved estimator \citep{Sidik2007}. \\
		``EB'' & Empirical Bayes estimator \citep{Morris1983}. \\
		``BM'' & Bayes modal estimator \citep{Chung2013}. \\ \hline
		`methodci' & Description \\ \hline
		NA & a confidence interval will not be calculated. \\
		``ML'' & Wald confidence interval with a ML estimator \citep{Biggerstaff1997}. \\ 
		``REML'' & Wald confidence interval with a REML estimator \citep{Biggerstaff1997}. \\ \hline
	\end{tabular}
\end{table}

The REML estimator for $\hat{\tau}^2_{R}$ and its confidence interval based on $\hat{\tau}^2_{R}$ can be calculated as
\begin{verbatim}
	tau2h(sbp$y, sbp$sigmak, method = "REML", methodci = "REML")
\end{verbatim}

\subsection{Datasets in the pimeta package} \label{sec:illustrations_pimeta}

The pimeta package includes the following three published random-effects meta-analyses.
\begin{enumerate}
	\item Set-shifting data: The set-shifting data \cite{Roberts2007} included 14 studies evaluating the set-shifting ability in people with eating disorders.
	Standardized mean differences in the time taken to complete Trail Making Test between subjects with eating disorders and healthy controls were collected.
	Positive estimates indicate impairment in set shifting ability in people with eating disorders.
	\item Pain data: The pain data \cite{Hauser2009} included 22 studies comparing the treatment effect of antidepressants on reducing pain in patients with fibromyalgia syndrome.
	The treatment effects were summarized using standardized mean differences on a visual analog scale for pain between the antidepressant group and control group.
	Negative estimates indicate the reduction of pain in the antidepressant group.
	\item Hypertension data: The hypertension data \cite{Wang2005} included 7 studies comparing the treatment effect of anti-hypertensive treatment versus control on reducing diastolic blood pressure (DBP) in patients with hypertension.
	Negative estimates indicate the reduction of DBP in the anti-hypertensive treatment group.
\end{enumerate}
These datasets are available in the pimeta package by using the following code.
\begin{verbatim}
	data(setshift, package = "pimeta")
	data(pain, package = "pimeta")
	data(hyp, package = "pimeta")
\end{verbatim}
We applied the methods were introduced in Section \ref{sec:prediction} using the `pima()' function.
\begin{verbatim}
	pima(setshift$y, setshift$sigmak, seed = 2718281)
	pima(pain$y, pain$sigmak, seed = 1732050)
	pima(hyp$y, hyp$se, seed = 1414213)
\end{verbatim}
Results are summarized in Table \ref{tab:illust}.

%% Table 6
\begin{table}[h]
	\centering
	\caption{\label{tab:illust} Estimated results of the three published meta-analysis datasets in the pimeta package: the average treatment effect ($\hat{\mu}$) and its 95\% prediction intervals (PI: prediction intervals, HTS: the Higgins et al.'s prediction interval, PR-APX, PR-HK, PR-SJ, PR-KR: the Partlett--Riley's prediction interval with the approximate, Hartung--Knapp, Sidik--Jonkman, or Kenward--Roger's variance estimator, NNF: the Nagashima et al.'s prediction interval), and heterogeneity measures ($\hat{\tau}^2_{DL}$, $\hat{\tau}^2_{R}$, $I^2_{DL}$, and $I^2_{R}$).}
	\begin{tabular}{ccccc} \hline
		&  & Set-shifting & Pain & Hypertension \\
		& Method & $K=14$ & $K=22$ & $K=7$ \\ \hline
		95\%PI & HTS & 0.36 [$-$0.02, 0.74] & $-$0.43 [$-$0.84, $-$0.02] & $-$8.83 [$-$11.21, $-$6.46] \\
		& PR-APX & 0.36 [\phantom{$-$}0.06, 0.67] & $-$0.42 [$-$0.77, $-$0.07] & $-$8.92 [$-$12.74, $-$5.10] \\
		& PR-HK & 0.36 [\phantom{$-$}0.05, 0.68] & $-$0.42 [$-$0.78, $-$0.06] & $-$8.92 [$-$12.77, $-$5.07] \\
		& PR-SJ & 0.36 [\phantom{$-$}0.06, 0.67] & $-$0.42 [$-$0.77, $-$0.07] & $-$8.92 [$-$12.73, $-$5.12] \\
		& PR-KR & 0.36 [\phantom{$-$}0.04, 0.68] & $-$0.42 [$-$0.79, $-$0.06] & $-$8.92 [$-$15.98, $-$1.86] \\
		& NNF & 0.36 [$-$0.12, 0.85] & $-$0.43 [$-$0.91, \phantom{$-$}0.03] & $-$8.83 [$-$12.76, $-$5.51] \\ \hline
		$\hat{\tau}^2_{DL}$ &  & 0.023 & 0.034 & 0.639 \\
		$\hat{\tau}^2_{R}$ &  & 0.013 & 0.025 & 1.729 \\
		$I^2_{DL}$ &  & 22.5 & 44.9 & 69.4 \\
		$I^2_{R}$ &  & 14.5 & 36.9 & 86.0 \\ \hline
	\end{tabular}
\end{table}

As shown in Table \ref{tab:illust}, the Nagashima et al.'s prediction intervals were substantially wider than the Higgins et al.'s and Partlett--Riley's prediction intervals for the set-shifting and pain data.
This result is consistent with findings of the simulation studies that revealed that the Higgins et al.'s and Partlett--Riley's prediction intervals may be too short in a setting with $I^2<50\%$ and $K<25$ \citep{Partlett2017,Nagashima2018}.
Next, the Partlett--Riley's prediction interval with Kenward--Roger's variance estimator was extremely wider than the others for the hypertension data.
This result is also consistent with the simulation studies.
The Partlett--Riley's prediction interval with Kenward--Roger's variance estimator may be overly conservative in a setting when there is a considerable difference between the within-study variances \citep{Partlett2017}.

\section{Conclusion} \label{sec:conclusion}
In this paper, we briefly summarize the recent developments in methods of prediction intervals for random-effects meta-analysis \cite{Higgins2009,Partlett2017,Nagashima2018} and provides readable examples for multiple types of data with simple code.
The pimeta package provides a function to compute a prediction interval and generate a forest plot for the estimated results.
The R code used to generate the results described in this paper are given in the Supporting Information.
The analysis is easily performed in R with a series of R packages.

\subsection*{Acknowledgements}
This work was supported by JSPS KAKENHI Grant Number 19K20229.

%------------------------------------------
%---------------- Bib ---------------------
\bibliographystyle{vancouver}
\bibliography{refs.bib}

\end{document}